\documentclass[ ]{copernicus2}

\usepackage{color}
\usepackage[T1]{fontenc}

\begin{document}

\title{Possible phase transition in  plasma mirror modes
}

{\author[1,3]{R. A. Treumann}
\author[2]{W. Baumjohann$^a$}
\affil[1]{International Space Science Institute, Bern, Switzerland}
\affil[2]{Space Research Institute, Austrian Academy of Sciences, Graz, Austria}
\affil[3]{Geophysics Department, Ludwig-Maximilians-University Munich, Germany\protect\\
Correspondence to: Wolfgang.Baumjohann@oeaw.ac.at}

}

\runningtitle{Mirror mode}

\runningauthor{R. A. Treumann \& Wolfgang Baumjohann}

\received{ }
\pubdiscuss{ } 
\revised{ }
\accepted{ }
\published{ }


\firstpage{1}

\maketitle

  

\noindent\textbf{Abstract}.-- 
{Mirror modes in collisionless high-temperature plasmas represent macroscopic high-temperature quasi-superconductors. We explicitly calculate the bouncing electron contribution to the ion-mode growth rate, diamagnetic surface current responsible for the Meissner effect, and the weak attracting electric field.  The mean electric field turns out to be negligible. Pairing is a second-order effect of minor importance. The physically important effect is the resonant interaction between bouncing electrons and the thermal ion-sound background. It is responsible for the mirror mode to evolve as a phase transition from normal to quasi-superconducting state.  } 

\section{Introduction: Mirror mode evolution}

In previous work we investigated the role of electrons trapped in a developed ion mirror mode, a rather complex problem. We showed that a subpopulation of mirror-mode trapped electrons develops an attracting potential in their wake and may contribute to a partial Meissner effect which expels a substantial fraction of the the magnetic field from the interior of the mirror bubble \citep{treumann2019}. The potential may also trap another electron in which case it  temporarily forms an electron pair at the high plasma temperature. However, as we will show below, pair production, though interesting in itself, is less important for the dynamics of the mirror mode even though we have made use of it in previous investigations. The important real effect is, independent of pairing, just the resonant interaction of the trapped electrons near their bouncing mirror point with the thermal ion-acoustic background. The latter is unavoidably present at the high plasma temperatures in the magnetosheath for example \citep{rodriguez1975}, thus inevitably leading to interaction with a substantial subgroup of electrons \citep{lund1996}. This has a number of important effects, in particular generation of very large temperature anisotropies and their effect on the evolution of the mirror mode in causing the Meissner effect.

The mirror mode  is one of the two mutually exclusive basic macroscopic mhd plasma modes which evolve in collisionless (ideally conducting) plasmas exhibiting pressure or temperature anisotropies \citep{chandrasekhar1961,hasegawa1969,hasegawa1975,krall1973,gary1993}. Its almost trivially obtained \emph{necessary} condition is the requirement on the pressure anisotropy 
\begin{equation}\label{eq-1}
A_i\equiv \frac{P_{i\perp}}{P_{i\|}}-1\,\gtrsim 0
\end{equation}
that the magnetically perpendicular pressure $P_{i\perp}=NT_{i\perp}$ exceeds the magnetically parallel pressure $P_{i\|}=NT_{i\|}$. Hence, its critical temperature is $T_{ci\perp}=T_{i\|}$. This means for the mirror mode that 
\begin{eqnarray}\label{eq-2}
T_{i\|}-T_{i\perp} &>& 0\qquad \mathrm{stability, ``normal"~state} \nonumber\\[-2ex]
&&\\[-2ex]
T_{i\|}-T_{i\perp} &<& 0\qquad \mathrm{instability, ``superconducting"~state}\nonumber
\end{eqnarray}
a condition which reminds at the superconducting phase transition in low-temperature solid state physics. Nevertheless, though apparently of much simpler non-quantum physics, it  is still barely completely understood  \citep{treumann2004} in spite of a wealth of publications which have been devoted to its elucidation \citep[e.g.][see also their reference lists]{kivelson1993,kivelson1996,baumjohann1996,treumann1997,constantinescu2002,constantinescu2003,pokhotelov2001,pokhotelov2002,pokhotelov2003,pokhotelov2005} and the discussion of various contributions to its evolution like plasma and field gradients, electron temperature, finite Larmor radius effects, non-thermal distribution functions, and more. Particular interest has been given also to the  electron anisotropy as the source of low frequency electron whistlers (Lion Roars) excited in mirror bubbles \citep[][cf. their reference list for earlier references on Lion Roars]{baumjohann1999} and were shown to be detectable and identifiable as caused by the \emph{resonant} electron anisotropy of the trapped electron component.  Further investigations of Lion Roars elucidated their different properties depending on the location in the mirror bubble \citep{tsurutani2011,yao2019,shoji2012,noreen2017,ahmadi2018,breuillard2018,lucek2001,lucek2005,czaykowska1998}. Finally it was shown that a genuine bulk non-resonant electron anisotropy can also generate its own mirror mode on the background of ion mirror modes \citep{noreen2017} and that these modes had already earlier been observed though had not been properly identified yet \citep{treumann2018a}. Subsequently it was demonstrated that single particle effects may play a role in the further evolution of the ion-mirror mode \citep{treumann2019} in close analogy to superconductivity. 

In the following we show that the ion-mirror mode, indeed to some extent, resembles superconductivity in high temperature plasma. The high (about infinite) conductivity is of course given at those temperatures on all scales below the diffusive scale. Thus in contrast to solid state physics of metals it poses no problem. Similarity to superconductivity is found in two effects, the generation of attractive potentials \citep{treumann2018} between electrons thus permitting for pairing, and a weak Meissner effect. Here we show that pairing and electric fields are of minor importance in the evolution of the ion mirror mode. The important effect is that the resonant interaction of mirror trapped electrons with ion-acoustic waves causes a large anisotropy and a classical phase transition from normal to quasi-superconducting state whose effect is diamagnetic thus resembling the Meissner effect and superconductivity. 

For perpendicular temperature below critical, i.e. for positive anisotropy $A_i>0$, the plasma may undergo instability when in addition the ambient magnetic field $\mathbf{B}$ drops below a characteristic threshold intensity  
\begin{equation}\label{eq-3}
B<B_c\approx \sqrt{2\mu_0NT_{i\perp}A_i}\,|\sin\theta|
\end{equation}
a condition which follows from the growth Eq. (\ref{eq-4}). It represents the \emph{sufficient} condition for growth. Hence, at $T_{i\perp}\gtrsim T_{i\|}$ the plasma may undergo a transition from a homogeneous to a diamagnetic state if the magnetic field is sufficiently weak. Since the plasma is ideally conducting under those high temperature conditions, appearance of diamagnetism indicates that the transition is from normal state $T_{i\perp}\leq T_{i\|}$ to quasi-superconductivity. The above necessary condition is sometimes inverted into a condition on the (arbitrary) angle of wave propagation $\theta=\tan^{-1}{k_\perp/k_\|}$ \citep{kivelson1993} but the relevant physics is contained in the field threshold, not the angle. The two conditions together resemble the zero temperature solid state physics thresholds for  superconductivity suggesting that the mirror instability is kind of a high-temperature macroscopic quasi-superconducting though not a quantum but a classical plasma state. 

This conclusion though not obvious at first glance is in fact not unreasonable. For at the high plasma temperature the classical conductivity is practically infinite on all collisional time scales, even on the anomalous quasilinear or any wave-particle or wave-wave interaction time scales. Growth rates of most instabilities are much faster than any diffusive scales, and growth lengths are much shorter than any diffusive lengths. However, most plasma instabilities do not satisfy the other condition of superconductivity, which is the magnetic Meissner effect as its identifier. This is just the case in the magnetic mirror instability which evolves into surprisingly large plasma bubbles which to a high degree, though not completely, are void of magnetic fields. Usually this is interpreted as the ion-mirror instability effect of a feed-back interaction between the interior and external regions of a mirror bubble simply assuming that the low energy ion component is trapped inside the bubble, while hot non-trapped plasma can be exchanged between magnetically connected bubbles on the same magnetic flux tube to blow up one of the bubbles on the expense of the others. This nonlocal process, though classically intuitive, has never been confirmed, as the decay of one mirror bubble on the expense of another one is difficult to observe. Nonlocal processes,  on the other hand, require information transport back and forth and thus need time. It is therefore interesting whether the evolution of mirror bubbles can be understood on the basis of local processes alone. 

The state of the observed bubbles poses another problem. Even though one can imagine that exchange between neighbouring bubbles takes place and is probably realistic to assume also in presence of local processes, pressure balance is readily achieved already at low amplitudes. Plasma exchange alone can hardly be taken as agent of the very large bubble amplitudes observed for instance in the magnetosheath where the magnetic amplitude drop reaches between 30\%-50\%. Amplitudes of this size are deep inside the extremely strong nonlinear plasma state. They require further growth far beyond quasilinear stability.  There is no linear or nonlinear theory available that could generate anything of this kind if not some violent external process forces the plasma locally into such an extreme state. Otherwise violent processes of pumping plasma into the growing bubbles are required.

Among the external processes capable of causing large decreases in magnetic field are, for instance, sudden local injections of cold plasma. This may happen when unobserved  very small meteorites evaporate inside the plasma. This is not unreasonable and may sometimes happen anywhere in near-Earth space. For them the environment  appears very diluted. A very  small meteorite would locally inject a large amount of quickly ionizing expanding and diluting matter which pushes the magnetic field away. The conditions for this to happen could easily be estimated. When occurring near the magnetopause the newly injected plasma would for a short time push even the whole magnetosphere far in, effects which are occasionally observed in space but have not yet been related to any such effects.

Mirror modes with their large amplitudes do not evolve into large solitary bubbles caused in this way. At the contrary they organize in chains elongated along the ambient magnetic field  providing the large scale plasma volume a typical mirror bubble texture. In this case one rather looks for local effects capable of causing further growth of the mirror instability. Some time ago we suggested \citep{treumann2018,treumann2019} that this can be obtained by a rather complicated chain of effects which may combine in anisotropic high temperature  plasma  resembling low-temperature superconductivity \citep{bardeen1957}, which is particularly interesting because superconductivity is not expected to work at high temperatures of the order of $T\gtrsim 10$ eV. Under those conditions the linear mirror instability drives the low-amplitude ion mirror mode which grows at a rate \citep[see, e.g.,][]{noreen2017}
\begin{equation}\label{eq-4}
\frac{\gamma}{\omega_{ci}}\approx \frac{k_\|\lambda_i}{1+A_i}\sqrt{\frac{\beta_\|}{\pi}}\Big[A_i+\sqrt{\frac{T_{e\perp}}{T_{i\perp}}}A_e-\frac{k^2}{k^2_\perp\beta_\perp}\Big]
\end{equation}
where $A_e$ is electron pressure anisotropy, $k_\|\ll k, \lambda_i=c/\omega_i$ ion inertial length, plasma-$\beta=2\mu_0N(T_i+T_e)/B_0^2$, and initially $A_e\approx0$. It grows on the expense of the ion pressure anisotropy until quasilinear stabilisation limits its magnetic amplitude to finite but still small values the order of at most a percent in pressure equilibrium and vanishing ion anisotropy $A_i\approx 0$. At this stage the mirror mode becomes susceptible to the anisotropic electron component with the latter being activated. Referring to the dynamics of the electron component this happens because any electron pressure anisotropy $A_e$ contributes as well to the above growth rate but is initially negligible as long as the ion-mirror mode grows. Once it stabilizes quasi-linearly and the ion anisotropy vanishes, electron anisotropy takes over. But, this is still a small effect that awaits amplification, which is provided by the microscopic electron dynamics. 

\section{Trapped resonant electron effects}
Mirror-trapped electrons perform a bounce motion. Close to their mirror points in the low amplitude bubble their parallel velocity is very low. When it becomes of the order of the ion sound speed fall into resonance with the always present thermal background ion sound noise. and an attractive  potential evolves in their wakes at a distance just outside their individual Debye spheres \citep{nambu1985,treumann2018}. This potential is capable of trapping another low-energy electron whose mirror point is closely located. Both electrons thus becoming locked to the ion-sound background noise and drop out of their bounce motion, propagating along the magnetic field together with the sound wave. Since they are close to their mirror point, their parallel speed has become that of the sound wave $v_\|=c_s$ with almost all their initial energy now in the perpendicular direction. Assume that the resonant electrons are simply thermal electrons. Then $v_\|\sim v_{e}$, where $v_e$ is the thermal speed. Already in this case the resonant electrons obey a very large anisotropy 
\begin{equation}\label{eq-5}
A_{pe}\approx \frac{T_e}{m_ec_s^2}-1\approx \frac{m_i}{m_e}\gg1
\end{equation}
Assuming that the the non-resonant electron component is isotropic and the number density of resonant electrons $N_p\ll N_0$ this yields for the total electron anisotropy
\begin{equation}\label{eq-6}
A_e\approx \frac{N_p}{N_0}\frac{T_e}{m_ec_s^2}\approx \frac{m_i}{m_e}\frac{N_p}{N_0}
\end{equation}
In fact there is no need for the electrons to be thermal. They might as well preferentially be from the near thermal tail, in which case their number would be less but their anisotropy would increase by the factor $\mathcal{E}_e/T_e>1$. This anisotropy continues to drive the mirror instability from quasilinear state with isotropic quasilinear $\beta$ until formation of large mirror bubbles. If quasilinear equilibrium is reached at $A_i\approx0 $ the electron driven ion-mirror growth rate is
\begin{equation}\label{eq-7}
\gamma_e\approx {k_\|V_A} \sqrt{\frac{\beta}{\pi}\frac{T_e}{T_0}}\Big[A_e-\sqrt{\frac{T_0}{T_e}}\frac{k^2}{k_\perp^2\beta}\Big]
\end{equation}
This immediately yields the threshold magnetic field for the electron-driven mirror mode
\begin{equation}\label{eq-8}
B<B_{ce}\equiv \Big[2\mu_0N_pT_0\frac{m_i}{m_e}\sqrt{\frac{T_e}{T_0}}\Big]^{1/2}|\sin\theta|
\end{equation}
This critical field $B_{ce}$ is very large. In the magnetosheath, for instance, one has $T_e\sim10 T_0$, and $T_0\sim 30$ eV. This gives $B_{ce}\approx 22\sqrt{N_p}|\sin\theta|$ nT. This amounts to $B_{ce}\gtrsim 10^2$ nT. The average magnetosheath field of say $B_{ql}\sim30$ nT is already below this threshold. When the ion mirror mod stabilizes quasilinearla its field drops even below this value though not susceptibly, just very few per cent. The above threshold for the resonant electrons is thus far above any quasilinear stable value of the ion mirror mode.  Hence the electron contribution to further evolution of the ion mirror instability has practically no threshold. It will evolve  immediately when the ion mirror mode stabilizes quasilinearly and provides an magnetic bottle for the electrons to bounce back and forth between their mirror points.   

On the other hand, if quasilinearity drives the ion-mirror instability towards marginality with $\beta A_i\sin^2\theta\approx 1$  there is anyway no threshold on the growth 
\begin{equation}\label{eq-9}
\frac{\gamma_e}{kV_A}\approx \frac{\beta\cos\theta\sin^2\theta}{1+\beta\sin^2\theta} \sqrt{\frac{\beta}{\pi}\frac{T_e}{T_0}}\,A_e
\end{equation}
and the instability will in both cases grow. In the latter case the instability can be more oblique than the pure ion mirror instability with $\cos\theta\gtrsim 0.2+1/10\beta$ for $\beta>1$ as is the case in the magnetosheath, or $\theta\lesssim 76^\circ$. This shows that the bubbles can become less elongated than in the linear case. 

Since, however, the sound wave transports the locked electron into the stronger magnetic field, not affecting its perpendicular dynamics, the magnetic moment $\mu=\mathcal{E}_\perp(s)/B(s)$ of the electrons remains conserved. This implies that the perpendicular energy of the electron increases. Resonance with ion sound heats the electrons in perpendicular direction. This happens for all electrons in resonance near their distributed mirror point when moving along the field into stronger magnetic field regions. At the contrary when an electron has bounced and resonates after turning around with another ion-sound wave propagating into the weaker field region, the retardation of the resonant electron in parallel direction implies that part of its perpendicular energy is pumped into the wave above its thermal level. It amplifies the background acoustic noise.   

\subsection{Perpendicular heating of resonant electrons} 
Though the extension of the instability beyond the quasilinear state is satisfactory, the main important point is the interaction between the bouncing electrons and the thermal  ion sound background. It perfectly resembles the interaction of Fermi electrons with the acoustic oscillations of the lattice in solid state physics. In that case the resulting formation of pairs  was the important macroscopic effect. It produced a substantial population of paired spin-compensated quasi-bosonic electrons which were capable of condensing in the lowest energy level near the Fermi surface thus becoming superfluid ans escaping interaction with the lattice. 

In our case something similar happens with a large number of electrons which get into resonance with ion-acoustic waves when their speed near the bouncing points becomes comparable to the phase speed of the waves. This will always happen under mirror conditions of this kind and is therefore a general property of mirror configurations where its  effect  is the generation of an attractive potential, in the first place however  locking of the resonant electrons to the ion sound wave. As explained, the resonance forces the electron to move with the sound wave deeper into the high magnetic field bottle neck than permitted by the bounce motion. Hence bounce symmetry is broken in this case. 

The interaction between electrons and an ion acoustic background thus has the  effect that it splits the  electron distribution into two populations. Near the bounce mirror points the resonant population moves locks to the ion acoustic wave, crosses the mirror point and moves farther into the high magnetic field of the  mirror bottle neck. Its parallel energy is  $\mathcal{E}_\|\sim m_ec_s^2/2$, its perpendicular energy is $m_ev_\perp^2/2\approx \mathcal{E}_0$ its total initial energy. In contrast, the nonresonant population becomes reflected and returns to the bottle center. Thus the bubble centre should be populated with nearly isotropic cool electrons while the bottles necks are populated with highly anisotropic $A_e\gg1$ electrons with their perpendicular energy still increasing due to the constancy of their magnetic moments $\mu$. 

The energy gain can be modelled for a mirror magnetic field $B(s)=B_0+\frac{1}{2}B''_0 s^2$ with $s$ the coordinate along the field measured from its minimum $B_0$ in the centre. Shifting to the mirror point $B_m(s_m)$ yields $B(s)=B(s_m)+\frac{1}{2}B''_0(s-s_m)^2$ and fixing the parallel energy of the electron to the sound speed $c_s$, the magnetic moment $\mu$ remains conserved because the motion at $c_s$ is slow compared with the gyration time of the electrons. Hence, with $\Delta s=s-s_m$ the gain in perpendicular energy is 
\begin{equation}\label{eq-10}
\Delta\mathcal{E}_\perp=\mu B'(s)\Delta s=\mathcal{E}_m(B''_0/B_m)(c_s\Delta t)^2
\end{equation}
 where $\mathcal{E}_m\approx \mathcal{E}_0$ is the total energy of the electron in the mirror point, and $\Delta t\sim 2\pi\ell/\omega_{ce}$ is the $\ell$-th multiple of the gyration of the electron it performs on the way as long as it remains trapped in resonance with the ion-sound wave. Hence, $\ell\gg 1$ is a comparably large number, and the magnetic moment remains conserved in spite of the violation of the bounce. The ultimate perpendicular energy the electron can attain in this process is then roughly roughly twice the original energy when reaching into the bottle neck
\begin{equation}\label{eq-11}
\mathcal{E}_e\approx \mathcal{E}_0\Big[1+\frac{B_{max}}{B_m(\mathcal{E}_0)}\Big] \lesssim 2\mathcal{E}_0
\end{equation}

\subsection{Surface currents and Meissner effect}
The Meissner effect depends on the generation of surface currents which screen the magnetic field from penetration. In mirror modes they decrease the internal magnetic field thus contributing to deepen the mirror bubble and causing a steep about exponential gradient at its boundary.

The perpendicular heating of a substantial number $N_p$ of electrons  increases the perpendicular pressure,  causing some more inflation of the bubble on the expense of the parallel pressure. Locked electrons which in flux tubes close to the bottle boundary experience the strong external field when their gyroradius increased. More important, they experience the plasma and field gradients. Trapped in the bottle boundary they perform a gradient drift along the boundary and form surface currents. The diamagnetic current flowing  in the external field $B_{ext}$  is 
\begin{equation}\label{eq-12}
\mathbf{J}_{dia}=\frac{\mathbf{B}_{ext}\times\nabla_\perp P_\perp}{B^2_{ext}}
\end{equation}
The pressure gradient points inward into the bubble when accounting for the large excess in perpendicular pressure. This current causes the meso-scale diamagnetism which suppresses the internal field and increases the external field. It is of the same nature as the boundary field in the Landau-Ginzburg mechanism of superconductivity \citep{landau1941,ginzburg1950}.  

Arguably the above diamagnetic current would to some extent be compensated by the curvature current
\begin{equation}\label{eq-13}
\mathbf{J}_{curv}=\frac{P_\perp-P_\|}{RB^2_{ext}}\mathbf{B}_{ext}\times\mathbf{n}
\end{equation}
where $\mathbf{n}$ is the external normal to the mirror bubble such that this current for $P_\perp>P_\|$ opposes the diamagnetic current. However, the total plasma pressure difference enters here. Moreover, the curvature radius is comparably large, and the contribution of curvature is therefore not substantial. 

Nevertheless the curvature current provides a limitation on the size if the bubble when $R$ becomes small. Then the curvature current may come up for the diamagnetism. In that case one finds a limit for bubble growth imposed not by the outer pressure balance but by the geometrical properties of the bubbles themselves. 

Equating both currents yields the  ultimate condition for stability of the bubble when the curvature current stops any further diamagnetism. Observations suggest that the field is not completely excluded but just to, $\sim 75\%$, implying an attenuation of the magnetic field $|\log (B/B_0)|\approx 0.3\approx R/a\lambda_i$, which implies that the curvature radius becomes approximately $R\approx3.5\lambda_i$. In the magnetosheath the ion-inertial length, which is also the ion skin depth, amounts to roughly $\lambda_i\approx 100$ km. Within a few $a=O(1)$ this seems approximately to account for the typical ultimate penetration depths of the magnetic field. Putting the gradient scale $|\nabla_\perp|\sim a\lambda_i^{-1}$. In terms of the ratio of curvature radius to gradient length across the boundary we obtain approximately 
\begin{equation}\label{eq-14}
\frac{R}{\lambda_i} \approx a+\frac{N_0P_{\perp0}}{N_pP_{\perp p}} 
\end{equation}
The last term is the product of initial to final resonant electrons density and perpendicular pressure ratios. Since $N_p<N_0$ and $P_{\perp p}>P_{\perp0}$ one can just conclude that the mirror bubbles will stabilize when the curvature radius drops to several times the ion inertial length. In that case the magnetic field from the external region can penetrate into the mirror bubble but is partly attenuated by the diamagnetic effect of the resonant electrons and their surface currents on the boundary of the mirror bubble. In this sense mirror bubbles are electrically charged.

In this sense the mirror mode, as we already suggested earlier \citep{treumann2019}, is an analogue to superconductivity. We repeat that this is not a complete analogy because we do not refer to any real quantum effects at the high temperatures of the plasma, and there is no lattice which fixes the ions. However, the inertia of the ions prevents them contributing to the resonance with the ion sound waves and therefore does not lead neither to attractive wake potential we found earlier, to pair formation, nor the drop out from bounce motion. All these effects are  due to the presence of electrons which fall into resonance, cease bouncing, cause a huge anisotropy and by this continue driving the mirror mode further by generation of diamagnetic currents which flow in the mirror boundary layer.

\subsection{Negligibility of average electric potential field}
Trapped electrons when resonating with the ion-acoustic background develop a weak attractive electric wake potential field just outside a Debye distance from the electron. This applies to all electrons which are capable of resonance. The Debye distance from the resonating electron is a  consequence of the resonance not a consequence of Debye screening. Hence, in a bath of ion-acoustic waves all electrons if capable of resonance could generate a weak local attractive potential in their wakes independent of the Debye screening. Since all these local potential drops are of same polarity, they may trap another electron but remain uncompensated. They are distributed over the entire mirror bubble in some way maximizing their occurrence probability around the distributed electron mirror points in the bubble volume. In the average they attribute a weak macroscale electric potential to the mirror mode.

In order to find the distribution of the attractive potentials we need the distribution of mirror points $s_m$ in the bottle. The location along the magnetic field is obtained from putting the parallel velocity zero. In a mirror configuration we have for the bouncing particle
\begin{eqnarray}\label{eq-15}
s(t,r)&=&\frac{v_{0\|}(r)}{\omega_b(r)}\sin\omega_b(r)t, \quad \omega_b(r)=\sqrt{\frac{\mu}{m_e}B''_0(r)}\\
t_m(r)&=&\frac{\pi}{2\omega_b(r)},\quad\quad\quad~ s_m(r)=s(t_m,r)=\frac{v_{0\|}(r)}{\omega_b(r)}\\
v_{0\|}(r)&=&\sqrt{\frac{2}{m_e}\big[\mathcal{E}_0(r)-\mu B_0(r)\big]}
\end{eqnarray}
We indicated that the bottle is rotationally symmetric by letting the quantities depend on radius $r$. These expressions can be combined to give the relation between the mirror point, coordinate and velocity along the magnetic field 
\begin{equation}\label{eq-16}
s_m^2=s^2(t)+v_\|^2(t)/\omega_b^2
\end{equation}
Assume we had an isotropic initial electron Boltzmann distribution in the central plane of the mirror bubble. Replacing the parallel velocity by the mirror point $s_m$ then gives the parallel Boltzmann factor which describes the distribution of mirror points along a magnetic flux tube
\begin{eqnarray}\label{eq-17}
f_\|(s_m)&\propto& \exp-\frac{m\omega_b^2[s_m^2-s^2(t)]}{2T_e}\\
&&\mathrm {where}\quad \quad 0< s_m^2-s^2<\frac{2}{m_e}\big(\mathcal{E}_0-\mu B_0\big)\nonumber
\end{eqnarray}
Since mirror points only exist for $s<s_m$ one finds that for any given mirror point $s_m$ there is a continuous distribution of mirror points at smaller $s$. This is true as long as we only consider the parallel distribution. However the full distribution further restricts the number of mirror points because for a given energy $\mathcal{E}$ the parallel velocity is limited by the available energy $\mathcal{E}_0$ as indicated by the upper bound on the right of the last expression. Within these limits a Boltzmann distribution of electrons  obeys a continuous distribution of mirror points along a given field line.  The mirror   point distribution along a flux tube suggests that many electrons  develop attractive wake potentials and provide the interior of the mirror bubble with a distributed electric potential. In this sense mirror bottles will be weakly electrically charged entities. We write the full distribution  as function of magnetic moments 
\begin{equation}\label{eq-18}
f(\mu,s_m)\propto \exp\Big[-\frac{\mu B_0(r)}{T_e}\Big(1+\frac{s_m^2B''_0(r)}{2B_0(r)}\Big)\Big]
\end{equation}
The second term is the distribution of mirror points combined with the magnetic moment. Integrating over $\mu$ yields the relative number density of mirror points as function of radius 
\begin{equation}\label{eq-19}
\frac{N_m(r)}{N}=\frac{B_0''(r)}{2B_0(r)}
\end{equation}

Calculation of the charging of the mirror bubble is a major effort, however. In previous work we determined the attractive potential for resonant electron. Restricting to one electron only, the potential is given by
\begin{equation}\label{eq-20}
\Phi(s,\rho,t)=-\frac{e}{2\sqrt{2}\pi\epsilon_0}\frac{c_s}{v_\|}\frac{e^{-\rho/\lambda_e}}{\lambda_e\sigma^2}\Big\{\sin|\sigma|-|\sigma|\cos|\sigma|\Big\}
\end{equation}
The requirement that the braced expression must be positive yields the condition $0<\sigma<\pi/2$ on $\sigma=(v_\| t-s)/\lambda_e$. This expression must  be applied to the bouncing electron near its turning point $s_m$. We then have $v_\| t\approx s_m$ and 
\begin{equation}\label{eq-21}
0<s_m-s<{\textstyle\frac{1}{2}}\pi\lambda_e
\end{equation}
for the location $s$ of the (maximum) attracting potential with respect to the mirror point. (We refrain from maximizing the potential because its extension is just of the order of the Debye length. Also its radial extension is of the same order as indicated by the exponential in $\rho$.) For the single electron this  becomes
\begin{equation}\label{eq-22}
\Phi(v_\|,\rho)\lesssim -\frac{\sqrt{2}\,e}{\pi^3\epsilon_0\lambda_e}\frac{c_s}{v_\|}\big(1-\frac{\pi}{2}\Delta\big)e^{-\rho/\lambda_e}
\end{equation}
where we have put $\sigma\approx\frac{1}{2}\pi-\Delta$ when expanding, with $\Delta\ll1$. If putting some numbers in here when referring to the magnetosheath, we have, with $c_s\sim v_\|$ at $s\sim s_m$ roughly that $|\Phi|\lesssim 10^{-9}$ V, which clearly is a weak potential but was not expected to be high. The local electric field is $|E|\sim 10^{-10}$ V/m which, however, is probably sufficient outside the Debye sphere to briefly trap another electron of about same speed as we proposed earlier. Though this is possible, potential generation is not the most important effect as we will immediately see. 

Let us briefly rewrite the above potential and calculate its average  inside a mirror bubble. We have
\begin{equation}\label{eq-23}
v_\|(t)=\omega_b\sqrt{s_m^2-s^2(t)}\lesssim\omega_b\sqrt{\pi \lambda_es_m}=\sqrt{\pi\omega_b\lambda_ev_{0\|}}
\end{equation}
using this in the relation between $s_m^2$ and $s^2(t)$ we have for the bouncing electrons $v_\|^2(t)/\omega_b^2<\pi\lambda_es_m$. With this we are in the position to replace the parallel velocity $v_\|(t)$ with its starting value $v_{0\|}$ in the expression for the potential. Mirror bottles are filled with roughly cylindrical regions of attractive potential of radius $\lambda_e$ and parallel extension of $\sim 3\lambda_e$. The ensemble average of the potential along the magnetic field is obtained when integrating with initial parallel Boltzmann distribution 
\begin{equation}\label{eq-24}
\big\langle\Phi(\rho)\big\rangle\lesssim-\frac{\Gamma(1/4)ec_s}{\epsilon_0\lambda_e
\sqrt{8\pi^7\omega_b\lambda_e}}\Big(\frac{m_e}{2T_{0e}}\Big)^{1/4}\frac{N_p}{N}\big(1-\frac{\pi}{2}\Delta\big)\,e^{-\rho/\lambda_e}
\end{equation}
Clearly this potential is weak. In the magnetosheath is amounts to just roughly $\lesssim10^{-9}N_p/N$ V which corresponds to the above estimate just multiplied with the density ratio which reduces the potential quite a bit. The volume of the average attractive potential is $V_p\approx 3\pi\lambda_e^3$. A fraction $N_p/N$ of electrons carries this potential in the mirror bottle along. So the total volume occupied by the potential is $V_\Phi\sim N_pV_p/N$. This gives an average attractive potential contained in the mirror bubble of volume $V_m\approx \pi R^2L_\|$
\begin{equation}\label{eq-25}
\big\langle\Phi\big\rangle_{tot}\lesssim-\frac{3\Gamma(1/4)}{2\sqrt{2\pi}}\frac{\lambda_e^3}{\pi^3L_\|R^2}\frac{e}{\epsilon_0\lambda_e}\frac{c_s}{\sqrt{\omega_b\lambda_e}}\Big(\frac{m_e}{2T_{0e}}\Big)^{1/4}\Big(\frac{N_p}{N}\Big)^2
\end{equation}
This average potential is very small. It exists theoretically but for all processes can be neglected. The expected result is that the total field generated in this way can be neglected. A mirror bubble is charged but only very weakly which makes the electric effect negligibly small.
 
\section{Landau-Ginzburg approach -- Phase transition}
The important effect of resonance between electrons and ion sound waves near the mirror points of the particles is neither the generation of the attracting potential, as this is very small, nor the possible trapping of electrons. It is the locking effect of the electrons when they remain in resonance with the unavoidably present thermal ion-acoustic wave background noise. This resonant interaction is the reason for the generation of the very large anisotropy in the resonant electron population, its effect on the ion mirror instability, and the generation of diamagnetic surface currents which cause the Meissner effect and partially expel the magnetic field from the bubble. Since about all sufficiently high energy electrons in the mirror bottle perform a bounce motion there will always be a relatively large number of electron which at least for a limited time are in resonance with the acoustic background noise. Therefore the anisotropy will be dominated b this population. Since electrons are indistinguishable the presence of this population implies that the mirror bottle at quasiliner stability always contains a resonant component of relative density $n=N_p/N<1$. They carry free energy $F(n)$ which is given by the classical Landau-Ginzburg expression 
\begin{eqnarray}\label{eq-26}
F(n)&=& F(0)+\int d\mathbf{x}\Big\{\frac{\hbar^2}{2m_e}\big|\big[\nabla-\frac{ie}{\hbar}\mathcal{A}\big]\big|^2n\nonumber\\
&&+\,\alpha n+ {\textstyle\frac{1}{2}}\lambda n^2+\frac{B^2}{2m_e}\Big\}\\ 
&& \mathrm{where}\quad\nabla= i\mathbf{p}/\hbar\nonumber
\end{eqnarray}
where $\mathcal{A}$ is the magnetic vector potential \citep[cf., e.g.,][]{binney1999}. Since $n$ results from interaction of the electron distribution in a complicated way with the background thermal noise, it is not a simple function. Electrons are charged, reacting to the complex wave fields via their bounce mediated electrical interaction. Shortcutting this one may assume that $n(\mathbf{x})=|\psi|^2$ is the squared modulus of a complex function $\psi$. This has a number of advantages.

Variation with respect to $\mathcal{A}$  and $\psi*$ immediately produces Maxwells equation $\nabla\times\mathbf{B}=\mu_0\mathbf{j}$ and the Landau Ginzburg equations
\begin{eqnarray}\label{eq-27}
-\frac{\hbar^2N}{2m_e}\Big[\nabla-\frac{ie}{\hbar}\mathcal{A}\Big]^2\psi(\mathbf{x}) +\alpha\psi(\mathbf{x})+\lambda|\psi(\mathbf{x})|^2\psi(\mathbf{x})&=&0\\
-\frac{ie\hbar}{2m_e}\Big[\psi^*\nabla\psi-\psi\nabla\psi^*\Big] -\frac{e^2}{m_e}|\psi|^2\mathcal{A}&=&\frac{\mathbf{j}}{N}
\end{eqnarray}
One may note that the quantum mechanical formulation is required at this stage by the definition of the generalized magnetic moment. However in the derivative term in the first equation one realizes that the quantum will cancel in the factor of the vector potential while the mixed term will vanish in the classical limit, and in the equation for the current the first term is the quantum current which vanishes in the classical limit. This reproduces  the London equation for the current with penetration length $\lambda^2_L=m_ec^2\epsilon_0/e^2N_p=\lambda_e^2/n>\lambda_e^2$. In the classical limit the gradient will  disappear. The classical current 
\begin{equation}\label{eq-28}
\mathbf{j}_{cl}=-\frac{n\mathcal{A}}{2\mu_0\lambda_e^2}
\end{equation}
is the diamagnetic screening current. Since the penetration depth in fully developed real mirror mode bubbles is larger than the electron inertial length, the normalized density $n<1$ must be relatively small. 

Taking the classical limit in the first equation yields that
\begin{equation}\label{eq-29}
\psi\Big[\lambda n+\alpha+\frac{e^2N}{2m_e}\mathcal{A}^2\Big]=0
\end{equation}
which has two solutions $\psi=0$, which is the normal state where there are no electrons in state $p$ and thus $N_p=0$. The second solution is however
\begin{equation}\label{eq-30}
n=|\psi|^2=-\frac{1}{\lambda}\Big[\alpha+\frac{e^2N}{2m_e}\mathcal{A}^2\Big], \quad \lambda>0
\end{equation}
giving the number normalized density of the electrons which contribute to the Meissner effect. This is the expectation value of the field $\psi$ which must be real and positive. Hence  one has $\alpha =-\bar\alpha$ and thus
\begin{equation}\label{eq-31}
-\alpha=\bar\alpha>{\textstyle\frac{1}{2}}\,\epsilon_0\omega_e^2\mathcal{A}^2 =\frac{\mathcal{A}^2}{2\mu_0\lambda_e^2}
\end{equation}
The complex field $\psi = |\psi|e^{i\delta}$ has an undetermined phase $\delta$ whose range is $0\leq\delta\leq2\pi$. This implies, that in the direction $|\psi|$ the symmetry of the field is spontaneously broken when the condition on $\bar\alpha$ holds, while rotational symmetry is maintained in the direction $\delta$. Hence in the mirror mode we have the case of a classical breaking of gauge symmetry. 

The latter condition on the sign of $\alpha$ is intriguing in as far as it immediately suggests that $\alpha$ should be proportional to the basic necessary condition for the evolution of the mirror mode,  the pressure (temperature) difference or anisotropy. One may set after quasilinear stabilization 
\begin{equation}\label{eq-32}
\alpha= a'\big(T_\|-T_\perp\big)=-\bar{a}T_\|A_e, \quad \bar\alpha'=\bar{a}T_\|A_e
\end{equation}
noting that $T_{c\perp}=T_\|$ is the critical temperature for the mirror mode. Once $A>0$ and this condition is satisfied one in addition needs the condition on the critical magnetic field  $b=B_c\big(1-B_{mm}/B_c\big)>0$, where now $B_{mm}\approx \frac{1}{2}B_0$, is the maximum mirror mode amplitude. Hence, combining both conditions yields 
\begin{equation}\label{eq-33}
\bar\alpha=\bar\alpha'b = \bar{a}T_\|B_c\big(1-B_{mm}/B_c\big)A_e\approx \bar{a}T_\|B_c\big(1-{\textstyle\frac{1}{2}}B_0/B_c\big)A_e
\end{equation}
which is the necessary condition for the mirror mode under fully developed stationarity after phase transition has occurred.

If retaining the gradient term in classical theory one replaces it with the expression for the generalized momentum of the particles $\big(\mathbf{p}-e\mathcal{A}\big)$. In mirror modes the vector potential has only the component $\mathcal{A}_\theta$ while the momentum of the particles in resonance has two components $\mathbf{p}=\big(0,p_\perp,m_ec_s\big)$. With these substitutions the classical Landau-Ginzburg equation at broken symmetry becomes
\begin{equation}\label{eq-34}
n=|\psi|^2=-\frac{1}{\lambda}\Big[\alpha+\frac{Nm_ec_s^2}{2}+{N\langle\mathcal{E}_e\rangle}\Big(1-\frac{e\mathcal{A}_\theta}{\sqrt{2m_e\langle\mathcal{E}_e}\rangle}\Big)^2\Big]
\end{equation}
As before, $\alpha=-\bar\alpha$ must be negative. We are dealing with locked resonant electrons near their bounce-mirror point. The second term in the bracket is unimportant, while the ensemble averaged perpendicular gyration energy of the locked resonant electrons is $\langle m_ev_\perp^2/2\rangle\gtrsim \langle\mathcal{E}\rangle=n\mathcal{E}_e\approx 2n\mathcal{E}_0$. The vector potential in the final state is $\mathcal{A}_\theta\approx RB_{mm}\approx\frac{1}{2}B_0R$. Here $R\ll\lambda_\|^{-1}$ is the above used ultimate curvature radius. The condition for the quasi-superconducting mirror phase transition thus becomes that
\begin{equation}\label{eq-35}
\frac{\bar\alpha}{N} >\langle\mathcal{E}_e\rangle\Big(1-\frac{eRB_0}{4\sqrt{m_en\mathcal{E}_0}}\Big)^2\approx 2n\mathcal{E}_0, \quad n<1 
\end{equation}
which can be combined with Eq. (\ref{eq-33}) to obtain a condition on $\bar{a}$. Otherwise, up to the numerical factor $\bar{a}$, which can be determined from experimental conditions, this expression takes care of the final physical properties of the mirror mode in its stationary state after phase transition. It still depends on the vector potential which can be expressed through the London current $j_{cl}$ Eq. (\ref{eq-28}). This last expression, up to the undetermined factor $\lambda$  relates the fractional density  of locked electrons to the total average initial energy. 

Mirror modes thus turn out to be high temperature phase transitions from normal even anisotropic plasma state to a state where they to some extent resemble superconductors. Since the conductivity is already about infinite for all processes the decisive property is the diamagnetic one which the mirror mode attains in this process. It occurs on a relatively short plasma scale of the order of the London depth and provides the plasma with a particular texture which, of course, in its own vain may serve as an energy source for further evolution.

\section{Summary}
The mirror mode ist one of the most interesting macroscale effects in high-temperature magnetized  plasmas obeying a perpendicular pressure anisotropy.  It is well-known that such anisotropies can be enforced on the plasma by interaction of a streaming plasma with an obstacle, for instance when bow shock waves are generated \citep{czaykowska1998,lucek2005,balogh2013}. This is the case in magnetosheaths \citep{lucek2001,lucek2005,yoon2007} of magnetized planets or other magnetized celestial bodies of all kinds, in particular also supernova remnants where one knows that large and small scale blast  and bow shock waves evolve in the interaction with the remnant interior matter and external interstellar gases. In all those cases mirror modes necessarily evolve and form chains of rather big magnetic field diluted bubbles in pressure equilibrium along the magnetic field which are in quite a different plasma state where the plasma is quasi-superconducting. They provide the matter a special magnetic bubble texture the effects of which are still barely understood. Here we have elucidated the mesoscale physics of mirror mode bubbles investigating the reasons for their evolution up to large amplitudes. The physics of this evolution is rather different from that of a normal linear and also nonlinear instability as the conditions for the Meissner effect have to be provided by the growth mechanism. Neither the linear nor the nonlinear instability is bale of doing this. They just serve as the initial state of growth.  There are no normal classical processes known which could produce the observed large amplitude mirror bubbles. Our theory is based on the quantum mechanical analogue of superconductivity in the non-microscopic description. We have, referring to the bounce motion of the linear mirror mode trapped electrons, shown that by interaction with the always present thermal ion-acoustic noise background electrons near their mirror points become locked to the resonant ion-acoustic wave, escape from bounce motion, gain additional perpendicular energy due to the transport of the wave-locked resonant electron far beyond its bounce mirror point, and evolve into an extreme anisotropic state. This theory is relatively easy to understand. The large locked electron anisotropy then enters the ion-mirror growth rate letting the mirror mode grow from quasilinear to larger amplitudes. In this process surface currents are generated which generate the wanted (partial, not a complete) Meissner effect. This process is, as we have demonstrated, based on Landau-Ginzburg theory) in fact a second order thermodynamic phase transition similar to superconductivity in metals. Ginzburg-Landau theory provides a convenient description of the transition, which we have provided the exact conditions for. So far we have proceeded until the determination of these phase transition conditions. We have expressed them through the initial conditions on the mirror mode and the final electron anisotropy. It is reasonable to end the investigation at this stage. We can, however, envisage a number of future processes to investigate further. The first is the mutual interaction of two mirror bubbles when encountering each other. It can be anticipated that this corresponds to a partial Josephson junction which should have interesting consequences as it, at the conditions given should provide information about the strength of the surface current. Otherwise, a most interesting case is concerned around the question what will happen, when a mirror bubble interacts with an antiparallel field. One expects that then reconnection will take place. However this kind of reconnection should be rather different from the ordinary one because two different phases of plasma come into contact locally. The immediate application of both cases is interaction at the magnetopause.

\section*{Acknowledgments}
 This work was part of a brief Visiting Scientist Programme at the International Space Science Institute Bern. RT acknowledges the interest of the ISSI directorate as well as the generous hospitality of the ISSI staff, in particular the assistance of the librarians Andrea Fischer and Irmela Schweitzer, and the Systems Administrator Saliba F. Saliba. We acknowledge valuable discussions with M. Leubner, R. Nakamura, and Z. V\"or\"os.

\end{document}